\documentclass[pre,amsmath,amssymb,aps,floatfix,nofootinbib]{revtex4}
\usepackage{dcolumn}
\usepackage{upgreek}
\usepackage{bm}
\usepackage{soul}
\usepackage[dvips]{epsfig}
\usepackage{graphicx}
\usepackage{xcolor}
\usepackage{amsmath} \usepackage{amssymb}

\newcommand{\comment}[1]{}

\newcommand{\BEA}{\begin{eqnarray}}
\newcommand{\EEA}{\end{eqnarray}}

\newcommand{\bq}{\begin{equation}}
\newcommand{\eq}{\end{equation}}
\newcommand{\be}{\begin{eqnarray}}
\newcommand{\ee}{\end{eqnarray}}
\newcommand{\ba}{\begin{align}}
\newcommand{\ea}{\end{align}}

\newcommand{\A}{ A} 
\newcommand{\B}{ B} 
\newcommand{\C}{ C}

\renewcommand{\aa}{ \bar{a}} 
\newcommand{\bb}{ \bar{b}} 
\newcommand{\cc}{ \bar{c}}

\begin{document}
\title{The most likely common cause}

\author{ A. Hovhannisyan$^{1,2}$ and A. E. Allahverdyan$^{1}$ }
\affiliation{ 
$^{1}$Alikhanian National Laboratory (Yerevan Physics Institute), Yerevan, Armenia\\
$^{2}$Institute of Applied Problems of Physics, Yerevan, Armenia }

\begin{abstract}
The common cause principle for two random variables $A$ and $B$ is examined in the case of causal insufficiency, when their common cause $C$ is known to exist, but only the joint probability of $A$ and $B$ is observed. As a result, $C$ cannot be uniquely identified (the latent confounder problem). We show that the generalized maximum likelihood method can be applied to this situation and allows identification of $C$ that is consistent with the common cause principle. It closely relates to the maximum entropy principle. Investigation of the two binary symmetric variables reveals a non-analytic  behavior of conditional probabilities reminiscent of a second-order phase transition. This occurs during the transition from correlation to anti-correlation in the observed probability distribution.  The relation between the generalized likelihood approach and alternative methods, such as predictive likelihood and the minimum common entropy, is discussed.  The consideration of the common cause for three observed variables (and one hidden cause) uncovers causal structures that defy representation through directed acyclic graphs with the Markov condition.
\end{abstract}

\maketitle

\section{Introduction}
\label{intro}

Cause-and-effect analysis traditionally proceeds in deterministic set-ups \cite{hesslow}. Probabilistic causality was developed in Refs.~\cite{reich,suppes}, where the common cause principle (CCP) was formulated: given two correlated events that do not influence each other, look for a past event that determined (caused) their joint distribution; see \cite{billi,berko,szabo} for reviews. The historical development of CCP is traced out in \cite{szabo,mazz,sterg} and Appendix \ref{ccappdx}; see Appendix \ref{examples} for examples  of applications of the principle. We follow to its formulation via two dependent random variables (not events) \cite{berko,szabo}. Denote them by $\A=\{a\}$ and $\B=\{b\}$, where $\{a\}$ is a set of $|\A|$ values of $\A$. Now given that $\A$ and $\B$ do not influence each other, a necessary condition for the random variable $\C=\{c\}$ to be their common cause is
\BEA
p(a,b|c)=p(a|c)p(b|c),\quad {\rm for~any~}(a,b,c),
\label{cccp}
\EEA
where we denote for probability ${\rm Pr}[\A=a,\B=b,\C=c]=p(a,b,c)$. Here $\A$, $\B$ and $\C$ occur at times $t_\A$, $t_\B$ and $t_\C$, respectively. Eq.~(\ref{cccp}) assumes $t_\C<t_\A$ and $t_\C<t_\B$ \cite{suppes}. Within CCP causality is not defined independently. Thus, we find it indispensable to make possibly explicit the time-dependencies of the random variables and postulate that only earlier variables can cause later ones \cite{suppes} \footnote{This is consistent with relativity theory \cite{penrose,wharton}. We understand that time-labels of random variables may not be available in practice. }.

According to CCP correlations between $\A$ and $\B$ are explained via the lack of control over $\C$. One implication of (\ref{cccp}) is worth noting, since it implies that the common cause is a better predictor \cite{suppes}. Assume that $t_\B<t_\A$ and that $p(b,c)>0$ for all $(b,c)$. Note from (\ref{cccp}) that there exist values $c$ and $c'$ of $C$ such that [see Appendix~\ref{derivation}] 
\BEA
p(a|b,c')=p(a|c')\geq p(a|b)\geq p(a|b,c)=p(a|c).
\label{govinda}
\EEA
We propose that predictionally efficient causes will satisfy the inequalities in (\ref{govinda}) strictly.
Hence if $b$ prevents $a$ (i.e., $p(a)>p(a|b)$), then $c$ prevents better than $b$, while if $b$ enables $a$ (i.e., $p(a)<p(a|b)$), then $c'$ enables $a$ better than $b$. Altogether, (\ref{govinda}) with strict inequalities means that $C$ is a better predictor for $A$ than $B$, an important operational feature of the common cause. 

CCP is a methodological principle, but it can also be derived from certain premises \cite{berko,penrose,wharton}; e.g. assume that 
\BEA
\label{phiphi}
\A=\varphi(\C,\lambda), \quad \B=\phi(\C,\mu), \quad p(\lambda,\mu)=p(\lambda)p(\mu),
\EEA
where $\varphi(.,.)$ and $\phi(.,.)$ are certain functions, while $\lambda$ and $\mu$ are independent random variables (noises). Then (\ref{cccp}) follows \cite{berko}. Another derivation of CCP proceeds via the maximum entropy principle \cite{balian}, when the probability $p(a,b,c)$ in (\ref{cccp}) is recovered from maximizing the entropy $-\sum_{a,b,c}p(a,b,c)\ln p(a,b,c)$, assuming that only $p(a,c)=\sum_bp(a,b,c)$ and $p(b,c)=\sum_ap(a,b,c)$ are known and are to be imposed as constraints in the maximization.

CCP invites several open problems, e.g. which stochastic dependencies are to be explained via (\ref{cccp}), and when such explanations can be useful \cite{sober,hoover}? Our paper does not address this question, but rather its antipode: provided that we know that the common cause $\C$ exists, and given the joint probabilities $p(a,b)$ only (this situation refers to causal insufficiency, and $\C$ is also known as latent confounder \cite{scheines}), can we determine the most likely probabilities $p(a,c)$ and $p(b|c)$ that are constrained by (\ref{cccp})? Eq.~(\ref{cccp}) do not suffice for unique determination of $C$, i.e., (\ref{cccp}) defines a non-identifiable mixture model; see Appendix \ref{obs_non} for a brief reminder on maximum likelihood and nonidentifiability. Based on the generalized likelihood \cite{armen2020}|which for the present case closely relates to the maximum entropy principle|we show that the above question has a well-defined answer that is consistent with CCP. See Ref.~\cite{predo} and especially chapter 1 of Ref.~\cite{pawitan} for an in-depth review of various concepts and forms of likelihood.

After defining the concept of the most likely common cause in section \ref{gl}, we study in detail the simplest case of two binary, symmetric $\A$ and $\B$; see section \ref{bin}. Here we find that probabilities for the most likely cause demonstrate a non-analytic behavior (akin to a second-order phase-transition), when $p(a,b)$ changes from correlated to anti-correlated behavior. 
Section \ref{three} discusses the common cause for three variables. Here we find that the most likely cause can generate causal structures that cannot be represented via directed acyclic graphs with Markov condition.  Section \ref{beta>1} relates our generalized likelihood approach with predictive likelihood \cite{predo} and the minimal common cause entropy \cite{murad} which are alternative methods for recovering the hidden cause (or latent confounder). In contrast to the common cause inferred via the generalized likelihood, these methods offer two different types of sparse common cause, where some probabilities are set to zero. We argue that such sparse common causes demand prior information and that even if such information is available, the prediction by the predictive likelihood is more consistent with CCP than the minimal common cause entropy. We conclude in the last section.

Latent confounders (hidden causes) and their determination were already addressed in the literature. Ref.~\cite{jan1} discusses an inference problem that aims to determine whether statistical correlations between $A$ and $B$ are due to common cause or a direct cause from e.g. $A$ to $B$. The same problem is studied in Ref.~\cite{mdl} in a more general setting and via tools of statistics (minimum description length); see also \cite{jan2} in this context. Ref.~\cite{spek} assumes that besides the observed probabilities of binary $A$ and $B$, the functional models are known that generate them. These models amount to e.g. functions $\phi$ and $\varphi$ in (\ref{phiphi}), together with the corresponding noise models. For inferring latent confounders, Ref.~\cite{spek} applies methods and ideas of algebraic geometry.  Ref.~\cite{murad} addresses a problem related to ours, but aims at causal discovery, e.g. it offers an approach for distinguishing between a common cause versus a direct causation between the target variables $A$ and $B$. In the context of our present approach|which does assume the existence of a common cause as a prior assumption|we can compare with Ref.~\cite{murad} because once this prior assumption is accepted, Ref.~\cite{murad} does offer a method for inferring the common cause, which is based on the minimization of the common entropy.
Section \ref{beta>1} discusses this approach and its drawbacks and compares them with our results. Entropic and
information theoretic approaches to probabilistic causality are discussed in \cite{ay,ayay,dey,dey2,dom1,dom2}.

\section{Generalized likelihood}
\label{gl}

\subsection{Features of generalized likelihood}

Eq.~(\ref{cccp}) defines a mixture model with $\C$ being the hidden (latent) variable, and $\A\B$ being the observed part of the mixture. This is a non-identifiable mixture model, since for a given $p(a,b)$, (\ref{cccp}) does not determine the unknown parameters $p(a,c)$ and $p(b|c)$ of the mixture model uniquely, e.g. just count independent variables in (\ref{cccp}) \footnote{Denote by $|\A|$ the number of values (realizations) of $\A$. Write (\ref{cccp}) as (\ref{bvana}) and note that $|\C|(|\B|-1)$ free parameters come from the second condition in (\ref{bvana}), while $|\A|(|\C|-|\B|)$ come from the first condition in (\ref{bvana}). Altogether we have $|\C|(|\B|-1)+|\A|(|\C|-|\B|)$ free parameters. Even for the minimal case $|\A|=|\B|=|\C|=2$ this amounts to $2$ free parameters.}. Hence, the mixture model provided by (\ref{cccp}) is observationally non-identifiable. Nonetheless, the maximum likelihood principle can be generalized even for such mixture models \cite{armen2020}. For clarity, let us start with the mixture model $p_\theta(a,b,c)$, where $(a,b)$ are observed, $c$ is hidden, and $\theta$ is an unknown parameter. Define for this model the following generalized likelihood (GL) function:
\be
\label{beta0}
L_\beta(\theta)= \frac{1}{\beta} \sum_{a,b} p(a,b)\ln \sum_c p^\beta_\theta (a,b,c), 
\ee
where $0<\beta<1$ is a hyperparameter. Eq.~(\ref{beta0}) assumes a long sample of identical and independent values of $\A\B$ so that the mean of $\frac{1}{\beta}\ln \sum_c p^\beta_\theta (a,b,c)$ calculated on this sample leads to (\ref{beta0}); see Appendix \ref{obs_non}. It is seen that $L_1(\theta)$ is the marginal likelihood of the mixture model, but once we assume that the model is observationally non-identifiable, maximizing $L_1(\theta)$ over $\theta$ is useless because by assumption it has many maxima over $\theta$. 
As seen below, in the common cause case we encounter the extreme case of this degeneracy, where $L_1(\theta)$ does not depend on $\theta$ at all. 

Now maximizing $L_{\beta<1}(\theta)$ lifts this degeneracy even for $\beta\lesssim 1$, where one reason of taking $\beta\lesssim 1$ is that $L_{\beta\lesssim 1}(\theta)$ is close to $L_{1}(\theta)$. We emphasize that ${\rm argmax}[L_{\beta<1}]$ is generically a discontinuous function of $\beta$: its value for $\beta\to 1$ differs from ${\rm argmax}[L_{\beta=1}]$, because the maxima of $L_1$ are degenerate \footnote{Operating with $\beta\lesssim 1$ we pick up the non-trivial maxima of the generalized likelihood, and simultaneously we ensure that the features of such maxima do not strongly depend on $\beta$; i.e., the maxima of $L_{\beta=0.5}$ and $L_{\beta=0.95}$ do differ from each other, while the maxima of $L_{\beta=0.95}$ and $L_{\beta=0.99}$ do not differ much. The actual value of $\beta\lesssim 1$ is a matter of a practical trade-off since taking $\beta$ too close to $1$ will impede the numerical convergence to the non-trivial maxima. In practice, we observed that $\beta\in [0.95,0.99]$ performs sufficiently well. It is however not excluded that the trade-off values of $\beta$ will depend on the dimension of the studied problem. }  .

There are also fundamental reasons for employing $L_{\beta<1}(\theta)$ \cite{armen2020}. Firstly, it retains several general features of the usual likelihood. For example, $L_\beta(\theta)$ in equation (\ref{beta0}) is maximized over the unknown parameter $\theta$ and if we reparametrize $L_\beta$ via a bijective function $\psi(\theta)$, retaining full information on $\theta$, the maximization outcomes will be related through the same function: $\hat\psi = \psi(\hat \theta)$. Moreover, if $\theta$ lies within a partially convex set $\Omega$, where for $\theta_1, \theta_2 \in \Omega$ and $0< \lambda < 1$, there exists $\theta_3 \in \Omega$ such that $p_{\theta_3} = \lambda p_{\theta_1} + (1-\lambda) p_{\theta_2}$, then for $\beta \leq 1$, $L_\beta$ from (\ref{beta0}) is a concave function of $p_\theta$, since it is a linear combination of superposition of two strictly concave functions: $f(u) = u^\beta$ and $g(v) = \ln v$. Consequently, maximizing $L_{\beta<1}(\theta)$ over $\Omega$ yields a unique global maximum. 
Additionally, $L_\beta(\theta)$ adheres to the sufficiency principle, as discussed in \cite{armen2020}. In Appendix \ref{free-energy}, we explore another significant feature of the generalized likelihood (\ref{beta0}), specifically its relationship with free energy in statistical mechanics. A closely related aspect is that maximizing $L_{\beta\lesssim 1}(\theta)$ is equivalent to using the maximum entropy method, a well-known approach in probabilistic inference and statistical physics (see section \ref{maxent} and Refs.~\cite{balian, mdl, maxent_armen, lamont}). Moreover, maximizing $L_{\beta\to\infty}$ pertains to optimizing predictive likelihood (see section \ref{predosection} and Refs.~\cite{predo, pawitan}). Finally, as discussed below, the maximization of $L_\beta(\theta)$ has certain desirable features concerning the structure of the common cause principle.

\subsection{Application to the common cause principle}
\label{appccp}
To apply (\ref{beta0}) to our situation (\ref{cccp}), we note that $\theta$ just amounts to unknown probabilities $p(a,c)$ and $p(b|c)$:
\be
\label{beta1}
&& L_\beta= \frac{1}{\beta} \sum_{a,b} p(a,b)\ln \sum_c p^\beta (a,c) p^\beta (b|c),\\
&& p(a,b)=\sum_c p(a,c)p(b|c),\qquad 1=\sum_b p(b|c),
\label{bvana}
\ee
where we naturally also imply $0 \leq p(a,c)\leq 1$ and $0 \leq  p(b|c) \leq 1$. 
The generalized maximum likelihood method amounts to maximizing $L_{\beta<1}$ under constraints 
(\ref{bvana}); cf.~ (\ref{cccp}). The maximization result will generally depend on $\beta$, but we will see below 
that for $\beta$ there is a preferred range of values $\beta\lesssim 1$, and the result is nearly independent of $\beta$ in that range.
Note that $L_{\beta<1}$ is a concave function of $p(a,c)$ and $p(b|c)$, but constraints (\ref{bvana}) are quadratic over the unknown variables $p (a,c)$ and  $p (b|c)$, i.e., the maximization problem is not a convex optimization. Now $|C|$ (the number of realizations of $\C$) is also a fixed, and given {\it a priori} parameter during the maximization of $L_{\beta<1}$. 
 This assumption relates to a general feature of all maximum likelihood methods: they all produce a larger likelihood for a larger set of parameters. This leads to overfitting and a preference for complicated explanations over simpler ones, which is not desirable \cite{mathpsy}. In the context of identifiable models, special methods were designed (e.g. the minimum description length) to prevent this from happening, and to limit properly the parameter set of the maximum likelihood method; see \cite{mathpsy} for a review. The corresponding techniques for non-identifiable situations are in the process of development; see \cite{christensen,minent,mdl,maxent_armen}. Hence, we simply assume that $|C|$ is an {\it a priori} given parameter here.

\subsection{Consistency with the common cause principle}

The maximization of $L_{\beta<1}$ holds the following feature that is inherent in CCP: for independent $\A$ and $\B$, $p(a,b)=p(a)p(b)$, no cause is predicted whatsoever, i.e., the maximization of $L_{\beta<1}$ predicts:
\BEA
\label{ido}
p(a|c)=p(a), \quad p(b|c)=p(b), \quad p(c)=\frac{1}{|C|}.
\EEA
Eq.~(\ref{ido}) is deduced from the concavity of $L_{\beta<1}$; see Appendix \ref{anal1} for details. 


We now assume that $C$ is not given, while the joint probability $p(a,b)$ of $A$ and $B$ is known and given. We ask whether the 
generalized maximum likelihood method can be applied to those situations, where $A$ causes $B$ directly, or $B$ causes $A$ also directly.
To this end, note that (\ref{cccp}) is formally satisfied also with $\C=\A$ or with $\C=\B$. Now we find from (\ref{beta0}):
\BEA
\label{idol}
L_{\beta}(\C=\A)=L_{\beta}(\C=\B).
\EEA
It is seen that one cannot employ a partial maximization of $L_{\beta<1}$ for deciding whether $\A$ caused $\B$ directly or $\B$ caused $\A$. Put differently, if we were able to infer causality (between $A$ and $B$) then (\ref{idol}) would precisely not hold.

\subsection{Relations with the maximum entropy method}
\label{maxent}

The maximization of $L_{\beta \lesssim 1}$ relates to the maximum entropy method. Indeed, writing in (\ref{beta1}) $p^\beta (a,c) p^\beta (b|c)=p(a,c)p (b|c) e^{(\beta-1) \ln[p(a,c)p (b|c)]}$ and expanding over small $1-\beta$ we get
\be
L_{\beta\lesssim 1}&=& \sum_{a,b} p(a,b)\ln p(a,b) -(1-\beta)\sum_{a,b} p(a,c)p(b|c)\ln \sum_c p (a,c) p (b|c)+{\cal O}([1-\beta]^2)
\nonumber\\
&=& \sum_{a,b} p(a,b)\ln p(a,b) +(1-\beta)H(\A\B\C)+{\cal O}([1-\beta]^2),
\label{beta2}
\ee
i.e., the maximization of $L_\beta$ reduces to the maximization of the joint entropy $H(\A\B\C)=-\sum_{a,b} p(a,c)p(b|c)\ln \sum_c p (a,c) p (b|c)$ of $\A\B\C$ under constraints (\ref{bvana}). 
Since $H(\A\B)$ is given, this amounts to maximizing the conditional entropy $H(\C|\A\B)$
under constraints (\ref{bvana}). 

\section{Most likely minimal cause for symmetric binary variables}
\label{bin}

\subsection{Phase transition between correlated and anti-correlated situations}

Consider the simplest case of two binary random variables $\A=\{a,\aa\}$ and $\B=\{b,\bb\}$. Their joint probability is symmetric in the following sense:
\BEA
\label{pablo}
&& p(a,\bb)=p(\aa,b),\\
&& p(a)=p(b),
\label{picasso}
\EEA
where (\ref{picasso}) follows from (\ref{pablo}).
We carried out numerical maximization of $L_{\beta\lesssim 1}$ assuming in addition to (\ref{bvana}, \ref{pablo}) that the most likely cause $\C=\{c,\cc\}$ is binary. This produced the following result:
\begin{align}
\label{co1}
&{\rm for}\quad  p(a,b)p(\aa,\bb)>p^2(a,\bb):\quad
p(a|c)=p(b|c), \quad p(\aa|\cc)=p(\bb|\cc) ;\\
&{\rm for}\quad p(a,b)p(\aa,\bb)<p^2(a,\bb):\quad 
p(a|c)=p(b|\cc),\quad p(\aa|\cc)=p(\bb|c),\label{co2} \\ 
&\qquad \qquad \qquad \qquad \qquad \qquad \qquad 
p(c)=p(\cc).
\label{co22}
\end{align}
where $p(a,b)p(\aa,\bb)>p^2(a,\bb)$ in (\ref{co1}) means correlation. Likewise, $p(a,b)p(\aa,\bb)<p^2(a,\bb)$ means anti-correlation. 

One message of (\ref{co1}, \ref{co2}) is that symmetry (\ref{pablo}) of observed probabilities translates to symmetry relations (\ref{co1}, \ref{co2}, \ref{co22}) that involve the causing variable $C$. These relations are consistent with (\ref{beta1}, \ref{bvana}), as seen from writing them down for binary variables and employing (\ref{co1}). For the correlated situation $c$ ($\bar c$) acts on the same way on $a$ and $b$ (on $\bar a$ and $\bar b$). For the anti-correlated case $c$ ($\cc$) acts on $a$ ($\aa$) in the same way as $\cc$ ($c$) acts on $b$ ($\bb$). Since this intuitively obvious difference between correlated and anti-correlated has to respect (\ref{picasso}), we get (\ref{co22}), i.e., the common cause is unbiased in the anti-correlated situation. Note that generally $p(c)\not =p(\cc)$ for (\ref{co1}); cf.~Fig.~\ref{pac07}.

We can interchange $c$ by $\cc$ with corresponding changes in (\ref{co1}, \ref{co2}). This means that the maximum of $L_{\beta\lesssim 1}$ is degenerate. If this degeneracy needs to be fixed, we can assume as a prior information e.g. $p(b|c)\leq p(b|\cc)$. This fixation is needed if we vary parameters continuously and we want to deal with the same (continuously varying) cause. Fig.~\ref{pac07} shows an example of such a situation that can transit from anti-correlation to correlation due to changing $p(\aa,\bb)$. It is seen that in the vicinity of this transition, the derivatives of $p(a|c)$ and $p(a|\cc)$ (with respect to $p(\aa,\bb)$) are large by the absolute value and continuous. However, $p(b|c)$ and $p(b|\cc)$ vary continuously, but their first derivatives are discontinuous. Another quantity that does vary continuously, but with a discontinuous first derivative is $p(c)$ which is strictly equal to $1/2$ for the anti-correlated situation; cf.~(\ref{co2}, \ref{co22}) and Fig.~\ref{pac07}. 

Altogether, the situation resembles second-order phase transitions in statistical mechanics that are also accompanied by continuous, but non-smooth changes in the observed (order) parameters \cite{balian}. 
In statistical mechanics, the observed parameters are found from minimizing free energy, as the first derivatives of the free energy with respect to e.g. temperature or external fields \cite{balian}. For second-order phase transitions these parameters are continuous across phase transitions, but their derivatives over the control parameter are not. Altogether, the second derivatives of the free energy are discontinuous, hence the term ``second-order'' phase-transitions \cite{balian}. In our situation, the (observed) probabilities $p(a,c)$ and $p(b|c)$ [in (\ref{co1}, \ref{co2})] are found from maximizing entropy. Since they are continuous, but their first derivatives are not, we conclude that our situation is analogous to second-order phase transitions, rather than first-order phase transitions. 
However, in statistical mechanics, phase transitions take place only in the limit of infinitely many variables for the Gibbs probability that is also inferred via the maximum entropy under constraints that are linear over the sought probability \cite{balian}. For our situation, phase transitions are present already for three random variables. There is an interesting consequence of having a second-order (and not first-order) phase transition. It is known in statistical physics that during first-order phase transitions the two phases are basically independent of each other \cite{balian}. They are both locally stable and have a physical meaning above and below the phase-transition point. At this transition point, they interchange the global thermodynamic stability, as determined by the minimal free energy (the minimal free energy refers to the maximal likelihood). For second-order phase transitions a single phase is transformed through the transition point. Hence, the second-order transition means that the correlated and anti-correlated situations have the same (though transformed) most likely cause. A first-order transition would mean that the causes for these situations are different.

\subsection{Causation between events}
\label{ondat}

\begin{figure}[h!]
    \centering
    \includegraphics[scale=0.35]{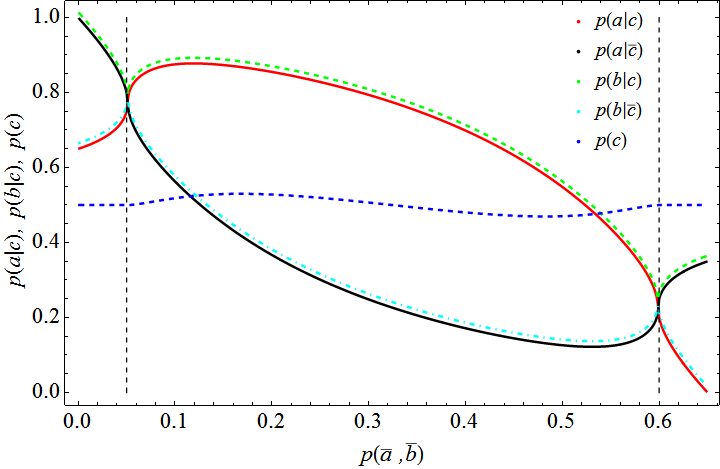}
    \caption{For three random variables $\A=\{a, \bar{a}\}$, $\B=\{b, \bar{b}\}$ and $\C=\{c,\bar{c}\}$, this figure presents conditional probabilities in (\ref{co1}, \ref{co2}) for a concrete numerical example $p(a,\bar{b})=p(\bar{a},b)=0.175$, where $0\leq  p(\bar{a},\bar{b}) \leq 0.65$. Thus $\A$ and $\B$ are correlated for $0.0511<  p(\bar{a},\bar{b}) <0.5989$ and are anti-correlated for $0< p(\bar{a},\bar{b}) <0.0511$ and $0.5989<  p(\bar{a},\bar{b})<0.65$; cf.~(\ref{co1}, \ref{co2}). Curves that are depicted close are in reality coinciding. The curve for $p(c)$ is seen to be non-monotonic with one global maximum and one global minimum. \\
    The maximization of $L_{\beta\lesssim 1}$ in (\ref{beta1}, \ref{bvana}) was carried out with $\beta = 0.95$ under additional condition $p(b|c)>p(b| \bar{c})$ that secures a continuous transition between correlated and anti-correlated situations. Several quantities are non-analytic along this transition, e.g. $p(b|c)$, $p(b| \bar{c})$, and $p(c)$. In particular, $p(c)=1/2$ in the anti-correlated situation. The maximization with $\beta =0.99$ produces the same results.   }
\label{pac07}
\end{figure}

Our results can be illustrated via the notion of event causality \cite{suppes,billi}. Here, earlier event $c$ causes later event $a$, if $p(a|c)>p(a)$, and among the relevant events, there is no event $f$ that can reverse this relation in the sense of $p(a|c,f)\leq p(a|f)$ \footnote{Note that even assuming $t_\B<t_\A$ we cannot apply this definition to an event $b$ causing $a$, because for $p(a|b)>p(a)$ we reverse this relation via $f=c$: $p(a|b,c)=p(a|c)$. }. 
This definition goes back to \cite{yule} and is reviewed in \cite{suppes,cox}. It can be applied to the above set of events with the conclusion that the relation [cf.~(\ref{co1}) and Fig.~\ref{pac07}]
\be
p(a|c)=p(b|c)>p(a|\cc)=p(b|\cc),
\ee
which holds for the correlated situation (\ref{co1}), means that $c$ causes (symmetrically) $a$ and $b$, while $\cc$ causes (symmetrically) $\aa$ and $\bb$. For the anti-correlated situation (\ref{co2}), $c$ causes $\aa$ and $b$, while $\cc$ causes $a$ and $\bb$; see Fig.~\ref{pac07}.

Finally, let us take a concrete example of (\ref{co1}):
\BEA
p(a,b)=0.1, \quad  p(a,\bb)= 0.01,\quad p(\aa,b)= 0.01,\quad p(\aa,\bb)= 0.88.
\label{lalo}
\EEA
This symmetric, correlated distribution [cf.~(\ref{pablo})] models the geyser example by Reichenbach \cite{reich}: two geysers ($A$ and $B$) are acting ($a$ and $b$) or passive ($\aa$ and $\bb$). Most of time they are passive together: $p(\aa,\bb)= 0.88$. They get active mostly also together, while there is a small probability $0.01$ of one acting without another. 

For (\ref{lalo}) we obtained numerically from maximizing $L_{\beta\lesssim 1}$ assuming $p(b|c)\geq p(b|\cc)$:
\BEA
\label{rr1}
&& p(c)=0.112974, \quad p(\cc)=0.887026,\\
\label{rr2}
&& p(a|c)=p(b|c)= 0.940956,\\ 
\label{rr3}
&& p(a|\cc)= p(b|\cc)= 0.004218,\\
&& p(\aa|c)= p(\bb|c)= 0.059443.
\label{rr4}
\EEA
The interpretation of (\ref{rr1}--\ref{rr4}) is as follows: the relatively rare event $c$ causes almost deterministically the activity of both geysers, but this is only a necessary condition for activity, not sufficient. Indeed, according to (\ref{rr4}) even if the causing event $c$ is there, the geysers need not act (presumably due to some local reasons). This explains why one geyser can be acting, while another one is passive.  The rough message of the geyser example is that rare correlated events most likely have a rare quasi-deterministic cause.

\section{Most likely common cause for three variables}
\label{three}

\subsection{Definition of the problem}

Let us now assume that $\A$ in (\ref{cccp}) consists of two random variables $\A=(A_1,A_2)$, where $A_1=\{a_1\}$ and $A_2=\{a_2\}$. Now we are given an observed probability $p(a_1,a_2,b)$ and look for the most likely common cause $\C$ for $\A$ and $\B$:
\BEA
p(a_1,a_2,b,c)=p(a_1,a_2,c)p(b|c),\quad t_{\C}<{\rm min}[t_{A_1}, t_{A_2}, t_{\B}],
\label{ccp3}
\EEA
where random variables refer to times $t_{A_1}$, $t_{A_2}$, $t_{\B}$, and $t_{\C}$, respectively. Relations between $t_{A_1}$, $t_{A_2}$, and $t_{\B}$ depend on concrete examples.

Note that joining two variables $A_1$ and $A_2$ into one $\A$ is not a trivial operation. It does imply that $A_1$ and $A_2$ are closer to each other than each of them to $\B$; e.g. $A_1$ and $A_2$ are contiguous in space or time or do belong to the same system. Here are two examples: {\it (i)} $A_1$ -- grades of examination, $A_2$ -- preliminary estimate of this grade, $\B$ -- name of a student arrived to examination, $\C$ -- knowledge gained before the examination; here $t_{A_1}>t_{\B}>t_{A_2}>t_{\C}$. {\it (ii)} $A_1$ -- cerebrovascular issues, $A_2$ -- cardiovascular issues, $\B$ -- susceptibility to stresses, $\C$-- genetic predisposition; now 
$t_{A_1}>t_{B}>t_{\C}$ and $t_{A_2}>t_{B}>t_{\C}$.

\subsection{Two extreme cases of initial 3-variable correlations}

We focus on two extreme cases of three-variable correlations in $p(a_1,a_2,b)$.
The first case when $A_2$ screens $A_1$ from $\B$:
\be
p(a_1,b|a_2)=p(a_1|a_2)p(b|a_2).
\label{guenter}
\ee
The second case when $A_1$ and $A_2$ are marginally independent:
\be
p(a_1,a_2,b)=p(b|a_1,a_2)p(a_1)p(a_2). 
\label{mah}
\ee
These two cases are representative in the following sense. Generic three-variable correlations in 
$p(a_1,a_2,b)$ can be quantified via the triple information $I(A_1:A_2:\B)$ that  can assume both negative and positive values (in contrast to the two-variable mutual information that is always non-negative); see Appendix \ref{tritri} for details. Moreover, three-variable correlations can be classified via the sign of $I(A_1:A_2:\B)$. Now (\ref{mah}) is a representative example of $I(A_1:A_2:\B)<0$, while (\ref{guenter}) is an example of $I(A_1:A_2:\B)>0$; cf.~Appendix \ref{tritri}.

\begin{figure}[h!]
    \centering
    \includegraphics[scale=0.35]{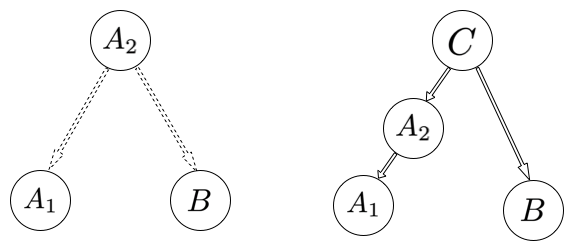}
    \caption{The left Causal Markov Conditioned Directed Acyclic Graph (DAG) denotes the correlation structure in (\ref{guenter}). The right DAG denotes the correlation structure implied by the most likely cause $\C$ of $\A=(A_1,A_2)$ and $\B$; see (\ref{uto}). 
    }
\label{fig2}
\end{figure}

\begin{figure}[h!]
    \centering
    \includegraphics[scale=0.35]{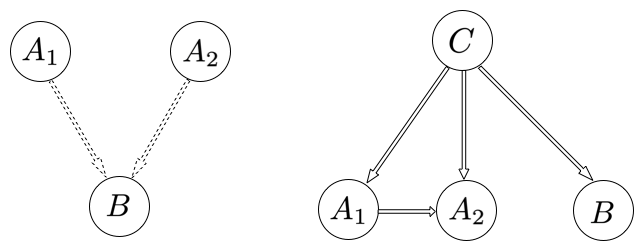}
    \caption{The left Causal Markov Conditioned Directed Acyclic Graph (DAG) denotes the correlation structure in (\ref{mah}). The right DAG denotes the correlation structure implied by the most likely cause $\C$ of $\A=(A_1,A_2)$ and $\B$.}
\label{fig3}
\end{figure}

Applying the maximization of $L_{\beta\lesssim 1}$ to (\ref{guenter}) we get
that the maximum is achieved for [see Appendix \ref{cond_ineq} for derivation]:
\be
\label{ut}
p(c|a_1,a_2)=p(c|a_2).
\ee
Supplementing (\ref{ut}) with the common cause definition (\ref{ccp3}) we find
for the most likely cause:
\be
\label{uto}
p(a_1,a_2,b,c)=p(a_1,a_2,c)p(b|c)=p(c)p(a_2|c)p(a_1|a_2)p(b|c),
\ee
which means that (\ref{uto}) can be depicted by the right graph in Fig.~\ref{fig2}; see (\ref{ut}).
Quantities in the right-hand-side of (\ref{uto}) are found from the maximization of $L_{\beta\lesssim 1}$, and do not have other general regularities. Note that (\ref{ut}) implies for mutual informations
\begin{equation}
\label{mut_ineq}
I(A_1: A_2 | \C)\equiv \sum_{a_1,a_2,c}p(a_1,a_2,c)\ln \frac{p(a_1,a_2|c)}{p(a_1|c)p(a_2|c)}=I(A_1:A_2)-I(A_1:C) < I(A_1:A_2),
\end{equation}
which means that not only global correlations are explained by the most likely cause (i.e., $I(A_1A_2:B|C)=0$), but also local correlations (i.e., the correlations between $A_1$ and $A_2$) are partially explained by the most likely cause $\C$. 

When looking for the most likely common cause of (\ref{mah}) one would expect to get $p(a_1,a_2|c)=p(a_1|c)p(a_2|c)$ in addition to the common cause definition (\ref{ccp3}).
This is however not the case \footnote{We can, of course, study $p(a_1,a_2,c)=p(c)p(a_1|c)p(a_2|c)$ via the generalized likelihood method, but only after imposing this structure as prior information, i.e., as an additional constraint during the maximization.}: the maximization of $L_{\beta\lesssim 1}$ produces (\ref{ccp3}), where $p(a_1,a_2,c)$ does not have any specific structure. In particular, we have now
\be
0=I(A_1:A_2)\leq I(A_1:A_2|\C).
\ee
We checked that a generic $p(a_1,a_2,c)$ is also obtained for two other non-trivial cases:
\be
p(a_1,a_2,b)=p(b)p(a_1|b)p(a_2|b),\\
p(a_1,a_2,b)=p(b)p(a_2)p(a_1|a_2,b).
\ee

\subsection{DAGs and TODAGs}

Let us now show how the obtained results can be interpreted in terms of Markov Conditioned Directed Acyclic Graphs \cite{scheines,berko}. We shall call them simply DAGs \footnote{Recall the definition of the Markov condition for DAG. Nodes of a DAG can be classified as parents and descendants. Now the Markov condition means that for any node $X$, ${\rm Pr}(X|{\rm parents~}X, {\rm non}-{\rm descendant~}X)={\rm Pr}(X|{\rm parents~}X)$; see e.g. \cite{scheines}. } and emphasize that any DAG can be employed in two different senses; see e.g. \cite{scheines,verma}. First, it can only reflect the correlation structure present; e.g. there are three DAGs that refer to (\ref{guenter}). One of them is the open fork shown in the left Fig.~\ref{fig2}, while another two of them are $A_1\to A_2\to\B$ and $\B\to A_2\to A_1$. The choice of DAG is reduced after it is postulated that the arrows of DAG coincide with the flow of time. For example, the unique DAG that refers to (\ref{guenter}) with condition $t_{A_1}<t_{A_2}<t_{\B}$ is $A_1\to A_2\to\B$. We shall call it time-ordered DAG (TODAG), whenever the emphasis is needed. But now it is possible that no TODAG exists for the given correlation structure; e.g. (\ref{mah}) cannot be represented via a TODAG, if $t_B<t_{A_1}$ or $t_B<t_{A_2}$; cf.~the left Fig.~\ref{fig3}. This is excluded from consideration as a non-generic situation \cite{scheines}. For this specific example, the exclusion is perhaps understandable, since it is empirically not expected to find two independent later variables influenced by an earlier one. 

Returning to (\ref{uto}), we see (given our basic assumption $t_{\C}<{\rm min}[t_{A_1}, t_{A_2}, t_{B}]$) that it can be described via a TODAG only for $t_{A_2}<t_{A_1}$. Whenever $t_{A_2}>t_{A_1}$ 
no TODAG exists that corresponds to (\ref{uto}). Note that the observed probability (\ref{guenter}) for $t_B>t_{A_2}>t_{A_1}$ refers to a well-defined TODAG: $A_1\to A_2\to B$. 

In contrast, $p(a_1,a_2,b,c)$ produced as the most likely cause from (\ref{mah}) does always have a unique TODAG, because $p(a_1,a_2,c)$ is generic, and we assume that $t_\C$ is the earliest time. This TODAG is unique even if for (\ref{mah}) no TODAG exists due to e.g. $t_B<t_{A_1}$. Thus, DAGs that refer to (\ref{guenter}, \ref{mah}) produce quite different most likely common causes. 

\section{Predictive likelihood, minimum entropy, and sparse common cause }
\label{beta>1}

\subsection{Predictive (maximum aposteriori) likelihood}
\label{predosection}

In section \ref{gl} we defined the generalized likelihood $L_\beta$ and argued that it is a concave function of $\beta$ for $\beta<1$; see (\ref{beta0}). Choosing $\beta\lesssim 1$ was motivated by this concavity, closeness to the marginal likelihood $L_{\beta=1}$, and relations with the maximum entropy method. We can also consider maximization of $L_{\beta> 1}$ motivating it as follows from (\ref{beta0}). Once in the joint probability $p_\theta(a,b,c)$ the hidden variable $c$ is unknown, we can estimate it via the maximum aposteriori method: given the observed $(a,b)$ we find 
\BEA
\label{dodo}
{\rm argmax}_c[p_\theta(a,b,c)]=\hat c(a,b,\theta). 
\EEA
Eq.~(\ref{dodo}) implies that some $c$ is kept fixed when generating each $(a,b)$, which may or may not hold true. Putting (\ref{dodo}) back into $p_\theta(a,b,c)$ we find unknown $\theta$ via the predictive likelihood \cite{predo}: 
\BEA
\hat \theta={\rm argmax}\Big[\sum_{a,b}p(a,b)\ln p_\theta(a,b,\hat c(a,b,\theta))\Big]={\rm argmax}\Big[L_{\beta\to\infty}\Big],
\label{otto}
\EEA
where we expressed $\hat \theta$ via $L_{\beta\to\infty}$ by noting from (\ref{beta0}): $\sum_c p^\beta_\theta (a,b,c)= p^\beta_\theta (a,b,\hat c(a,b,\theta))$, which is valid in the limit $\beta\to\infty$. Note from (\ref{beta1}, \ref{beta2}) that ${\rm max}[L_{\beta\gtrsim 1}]$ is equivalent to minimizing the full entropy $H(\A\B\C)$. 

Recall that $\theta$ amounts to unknown probabilities $p(a,c)$ and $p(b|c)$, as discussed in section \ref{appccp}. Now the maximization of $L_{\beta>1}$ is straightforward. We shall illustrate it for binary variables $\A=\{a,\aa\}$, $\B=\{b,\bb\}$, and $\C=\{c,\cc\}$. Take the first term in $L_{\beta>1}$ and note that
\BEA
p(a,b)\ln\Big(\Big[ p^\beta(a,c)p^\beta(b|c)+p^\beta(a,\cc)p^\beta(b|\cc)  \Big]^{1/\beta}\Big)
\leq 
p(a,b)\ln\Big( p(a,c)p(b|c)+p(a,\cc)p(b|\cc)  \Big )=
p(a,b)\ln p(a,b).
\label{buk}
\EEA
where we employed constraints (\ref{bvana}). 
Similar inequalities hold for every term in $L_{\beta>1}$. Hence, the maximization of $L_{\beta>1}$ is achieved for one of the following two solutions, where $\leq$ in (\ref{buk}) is satisfied as $=$:
\BEA
\label{ban1}
&&p(a,c)=p(\aa,\cc)=0,\\
&&p(b,c)=p(\bb,\cc)=0.
\label{ban2}
\EEA
They are equivalent since they provide the same maximal value ${\rm max}[L_{\beta>1}]=\sum_{k=a,\aa; l=b,\bb}p(k,l)$. Another two solutions are found from (\ref{ban1}, \ref{ban2}) by interchanging there $c$ with $\cc$; this symmetry is anyhow always present. Note that neither (\ref{ban1}, \ref{ban2}) nor ${\rm max}[L_{\beta>1}]$ depend on $\beta$ provided that $\beta>1$ \footnote{It is seen that (\ref{ban1}, \ref{ban2}) lead to sparse solutions, whereas ${\rm argmax}[L_{\beta<1}]$ does not contain sparse solutions. This feature makes an interesting analogy with statistical mechanics, where $\beta$ corresponds to the inverse temperature \cite{balian}. Physical systems with a sufficiently large $\beta$ (small temperatures) tend to show the ground-state dominance feature \cite{balian}, i.e., relatively small portions of their state space contribute to the observable quantities. This relates to the sparsity of the above probabilities.}. 

Each solution (\ref{ban1}) or (\ref{ban2}) is worked out using (\ref{bvana}); e.g. for   (\ref{ban1}) we get 
\BEA
p(a|\cc)=1, ~~ p(a|c)=0, ~~ p(c)=p(\aa), ~~ p(b|\cc)=p(b|a), ~~ p(\bb|c) = p(\bb|\aa). 
\label{alisa}
\EEA

\subsection{Minimum entropy}

For inferring the unknown common cause $C$ in (\ref{cccp}), Ref.~\cite{murad} proposed to minimize the entropy $H(\C)=-\sum_cp(c)\ln p(c)$ of $C$ \footnote{Recall, however, that the main aim of Ref.~\cite{murad} was in causal discovery and not in studying the common cause assuming that it is given.}. 
The rationale of this is that $C$ is made to be possibly deterministic: the absolute minimum $H(C)=0$ is reached for a deterministic $C$, where only one probability among $\{p(c)\}$ is non-zero. This minimal entropy is also known as the common entropy of $A$ and $B$. 

The concept of common entropy and related quantities (e.g. common information) is well established and plays a fundamental role in information theory; see \cite{chin} for a recent review. The common entropy for the considered binary case was studied in Ref.~\cite{kumar}, where also several general bounds on the common entropy minimizer $C$ were established. In particular, these bounds imply that whenever $A$ and $B$ are binary, for minimizing the common entropy it suffices to take a binary $\C=\{c,\cc\}$ \cite{kumar}. 
However, for our present purposes we find more instructive to minimize $H(\C)=-\sum_cp(c)\ln p(c)$ directly; see 
Appendix \ref{ff}. Thus, for binary $A$, $B$ and $\C=\{c,\cc\}$, the minimization of $H(\C)=-p(c)\ln p(c)-p(\cc)\ln p(\cc)$ under conditions (\ref{bvana}) produces two different candidate solutions (\ref{are1}) and (\ref{are2}), 
\BEA
\label{are1}
&&p(a|c)=p(b|c)=0,\\
\label{are2}
&&p(\aa|c)=p(\bb|c)=0,
\EEA
for the correlated situation \footnote{To show the first equality in (\ref{aumwelt}) note that $p(a)p(b)-p(\aa)p(\bb)=p(a)+p(b)-1$, and then expand $p(a)$ and $p(b)$ over the joint probabilities. }:
\BEA
p(a,b)-p(a)p(b)=p(\aa,\bb)-p(\aa)p(\bb)=p(a,b)p(\aa,\bb)-p(a,\bb)p(\aa,b)>0.
\label{aumwelt}
\EEA
The choice between (\ref{are1}) and (\ref{are2}) is done as follows. 
We work out (\ref{bvana}, \ref{are1}): 
\BEA
p(a|\cc)=p(a|b), ~~ p(b|\cc)=p(b|a), ~~ p(\cc)=p(a)p(b)/p(a,b),
\label{vox}
\EEA
and place $p(\cc)$ from (\ref{vox}) into $H(C)$. The same is repeated with (\ref{bvana}, \ref{are2}), and the two $H(C)$ are compared to each other. We take the candidate solution with the smallest $H(C)$ as the actual ${\rm argmin}\Big[H(C)\Big]$; see Appendix \ref{ff}. 
Note from (\ref{vox}) that weakly-correlated random variables, i.e., $p(a)p(b)\lesssim p(a,b)$, have vanishing common entropy. 

For the anticorrelated situation the inequality in (\ref{aumwelt}) is inverted. Now we get instead of (\ref{are1}) and (\ref{are2}) two different candidate solutions (\ref{are3}) and (\ref{are4}):
\BEA
\label{are3}
&&p(\aa|c)=p(b|c)=0,\\
\label{are4}
&&p(a|c)=p(\bb|c)=0.
\EEA
The choice between candidate solutions (\ref{are3}) and (\ref{are4}) is made following the same method as when choosing between (\ref{are1}) and (\ref{are2}): we work out (\ref{bvana}, \ref{are3}) and (\ref{bvana}, \ref{are4}), and select the solution with the smallest $H(C)$. 

We recall that the solutions (\ref{are1}--\ref{are4}) have their analogs with $c$ replaced by $\bar c$. They are equivalent to (\ref{are1}--\ref{are4}).  

\subsection{Defects of sparse common cause}

Eqs.~(\ref{ban1}, \ref{ban2}) and (\ref{are1}--\ref{are4}) are similar to each other in the following sense of sparsity: both assume the maximal number of zeros in $p(a,c)$ and $p(b|c)$ so that conditions (\ref{bvana}) provide a unique solution for given $\{p(a,b), p(a,\bb), p(\aa,b), p(\aa,\bb)\}$. 

Whether (\ref{ban1}, \ref{ban2}) or (\ref{are1}--\ref{are4}) should be preferred to the solution found via ${\rm max}[L_{\beta\lesssim 1}]$? In our opinion no, since (\ref{ban1}, \ref{ban2}) and (\ref{are1}--\ref{are4}) imply a mixture of deterministic and fine-tuned event causation [cf.~section \ref{ondat}] with probabilistic causation. Such a mixture is not realistic; i.e., it needs prior information. 

In addition, the common cause holding one of (\ref{are1}--\ref{are4}) has the following defect which is absent in (\ref{ban1}). Recall an important implication of the common cause principle|some common causes are better predictors|that demands $p(b,c)>0$ for all $(b,c)$; cf.~the discussion around (\ref{govinda}). This condition is violated both for each solution in (\ref{are1}--\ref{are4}); i.e., the common cause loses an important aspect of its meaning. In contrast, condition $p(b,c)>0$ holds for (\ref{ban1}). Depending on whether we want that $C$ is a better predictor than $B$ or $A$, we need to take (resp.) (\ref{ban1}) or (\ref{ban2}).  

To be more explicit, consider (\ref{vox}). Note that $p(a|\cc)=p(a|b)$ means that $\cc$ does not enable $a$ better than $b$, though $p(a|c)=0$ means that $c$ prevents $a$ better than $b$ prevents $a$. In contrast, looking at (\ref{alisa}) we note 
$p(a|\cc)=1>p(a|b)>p(a|c)=0$, which is the desired relation. 

Hence, even if a sparse common cause is searched for, we recommend the solution found via the predictive maximum likelihood, and not the one obtained via minimum entropy of $C$.

\section{Summary}\label{sum}

Using the generalized likelihood (GL) method, we determine the most probable unbiased cause $C=\{c\}$ for a given joint probability $p(a,b)$ of observed variables (effects) $A=\{a\}$ and $B=\{b\}$. This can serve as a reference point for causal inference, and also explains which causes are likely to be seen in practice. The choice to employ GL relates to the maximum entropy principle and is motivated by the non-identifiability of the problem. The results obtained from both numerical and analytical analyses affirm the effectiveness of the suggested approach. Notably, we observe an interesting phase transition-like behavior in the minimal setup when transitioning from anticorrelated to correlated $p(a, b)$. Furthermore, we present cases where the identified parameters can be interpreted in terms of time-ordered Directed Acyclic Graphs (TODAGs), as well as cases where such interpretation is not possible. We also compared our results with the minimum common cause entropy approach and identified disadvantages of that approach compared to predictions of GL and of the predictive likelihood. 

Several important questions are left for future research: {\it (i)} generalization of the GL method to causality with multiple variables. {\it (ii)} Variables with larger discrete domains. Here we note that maximum likelihood methods are known by computational feasibility, a subject that goes via the known Expectation-Maximization method.  It is shown in Ref.~\cite{armen2020} that the generalized likelihood also conforms to this method, i.e., we anticipate that for multi-dimensional discrete random variables, the calculations via the generalized likelihood will be feasible. {\it (iii)} Variables with continuous domains. Our preliminary results show that the concept of the most likely common cause generalizes well to (at least) Gaussian variables. {\it (iv)} The dependence of the GL function on the dimensionality of the common cause. Above we assumed that this dimensionality is known a priori. Elsewhere, we shall explore generalizations of the GL method, which omit this assumption. {\it (v)} Whether the GL method can distinguish the existence of a common cause {\it versus} the direct causal influence between the observed variables. {\it (vi)} Applications of the most likely common cause.

\begin{acknowledgments} This work was supported by SCS of
Armenia, grants No. 20TTAT-QTa003 and No. 21AG-1C038. 

\end{acknowledgments}

 \appendix

\section{Common cause principle and its evolution} 
\label{ccappdx}

Reichenbach \cite{reich} formulated the common cause principle with
respect to three distinct events $a$, $b$ and $c$; their joint probability
is $p(a,b,c)$. The definition amounts to the following 4
conditions \cite{reich,suppes,billi,berko,szabo,mazz,sterg}. 

\begin{enumerate}

\item The causal influence from $a$ to $b$ and from $b$ to $a$ are excluded. To this end, it is sufficient that the events a and b take place at equal times, or that they are space-like, i.e., no interaction could effectively relate $a$ and $b$. Naturally, $c$ refers to an earlier time, i.e., $c$ is in the common past of $a$ and $b$.

\item The correlation between $a$ and $b$ holds \cite{reich}:
    \BEA \label{corrs}
    p(ab) > p(a)p(b). 
    \EEA

    \item 
$c$ and $\cc$ screen-off correlations between $a$ and $b$:
\BEA \label{condind1}
    p(ab|c) &= p(a|c)p(b|c),\\
    \label{condind2}
    p(ab|\cc) &= p(a|\cc)p(b|\cc),
    \EEA
    where $\cc$ is the complement of $c$, i.e., the event $\cc$ takes place when $c$ does not and {\it vice versa}. 
    \item 
    $c$ increases the probability of $a$ $(b)$ compared to $\cc$
    \BEA \label{dcaus1}
    p(a|c) > p(a|\cc), \\
    p(b|c) > p(b|\cc).
    \label{dcaus2}
    \EEA
\end{enumerate}

The following objections were made against these conditions \cite{berko,uffink}.

--  It is not clear why specifically the correlation condition
(\ref{corrs}) is taken for defining correlations that need explanations;
anti-correlations also need explanations. 

-- If we accept (\ref{corrs}), conditions (\ref{dcaus1}, \ref{dcaus2}) are
redundant, because one can write using (\ref{condind1}, \ref{condind2}):
\BEA
        p(ab) - p(a)p(b) = p(c)p(\cc)[ p(a|c) - p(a|\cc) ][ p(b|c) -  p(b|\cc) ],
\EEA
i.e., if (\ref{corrs}) is accepted, then either $c$ or $\cc$ is a common
cause, there is no need to impose (\ref{dcaus1}, \ref{dcaus2}) separately. 

-- The existence of a single common cause ($c$ or $\cc$) is too
restrictive, so people moved to common cause systems, where instead of
two events with probabilities $p(c) + p(\cc) = 1$, one employs a larger
set of events $\{c\}$ with $\sum_c p(c) = 1$ \cite{szabo}.  Conditions
(\ref{condind1}, \ref{condind2}) generalize easily:
\BEA \label{multicondind}
        p(a,b|c) = p(a|c)p(b|c) \quad \text{for all}\quad  c.
\EEA
Note that conditions (\ref{dcaus1}, \ref{dcaus2}) do not extend uniquely
to this more general situation: there are at least two candidates for
such an extension and the choice between them is not unique
\cite{sterg,mazz}. 

 -- Reichenbach's formulation combines two different things: correlations that need to be explained via events $c$
and $\cc$, and event causation expressed by (\ref{dcaus1}, \ref{dcaus2}). The understanding of the event causation is blurred by Simpson’s and related paradoxes \cite{berko, billi}. Some of them question whether the cause need to increase the probability of its effect \cite{hesslow,beebee}. Hence, it is useful to separate explicitly the explanation of correlations from the event causation. It is also natural to avoid the restriction of doing only two specific values $a$ and $b$ for two random variables $A$ and $B$: we can require (\ref{multicondind}) for
all values of $A$ and $B$ under fixed $c$ \cite{uffink}. Once this is done, then we have to skip the condition
(\ref{corrs}) as well, because it cannot hold for all values of the
random variables $A$ and $B$. Summing
up, we converge to the definition given in section \ref{intro}.

\section{Examples of applications of the common cause principle} 
\label{examples}

Such examples frequently tend to be trivial or controversial. Interesting and non-controversial applications of the common cause to historical linguistics are given in \cite{lingvo}. 

The widely cited correlation of ice-crime sales and crime acts is curious, but not much more. The common cause here is the warm weather that leads to outdoor activities and tourist travel, thereby increasing crime. An old and important controversy relates to the known papers by Fisher \cite{fisher}, where he conjectured that the correlation data between smoking and lung cancer may have a genetic common cause. As of now, this conjecture is not supported. A more recent, and so far unresolved controversy is whether physical exercise causes longevity or there is a common cause that explains their observed correlation. There is no doubt that physical exercise improves overall well-being, but it may not affect longevity directly. Reasonable health practices, as well as the medical care system, can be the common cause here.

Cognitive science provides another interesting application of the common cause principle \cite{wegner}. People experience conscious willingness for some of their actions before taking them. Being willing to do something correlates with doing it. However, being willing something is most likely not the real cause of doing it \cite{wegner}. Several mechanisms show that both the action and the willingness to do it can be caused by unconscious mechanisms \cite{wegner}. More than being a direct cause of the action, willing the action before doing is probably connected with the need to accept responsibility for it and correct it in the future. 

\section{When the common cause is a better predictor } 
\label{derivation}

To deduce the first inequality in (\ref{govinda}) assume that $p(a|b)< p(a|b,c)=p(a|c)$ for all $c$, multiply both parts by $p(b,c)>0$, sum over $c$ and get contradiction $p(a|b)>p(a|b)$ \cite{suppes}. Note that to get the contradiction, it suffices to have $p(b,c)>0$ for one value of $c$. In the same way, the second inequality in (\ref{govinda}) is proved.

In the context of (\ref{govinda}) it is important to stress that if $p(a)<p(a|b)=1$, i.e., $b$ is the strongest cause of probabilistic event $a$, the common cause $C$ in the sense of (\ref{cccp}) does not exist provided that $p(b,c)>0$ for all $c$ \cite{suppes}. This is shown as follows, first let us assume that $1>p(a|bc)$, multiply both parts by $p(b,c)>0$, sum over all $c$, and get a contradiction with $p(a|b)=1$. Hence $1=p(a|bc)$. Now note that $p(a)=\sum_c p(a|c)p(c)=\sum_c p(a|bc)p(c)=\sum_c p(c)=1$, which is a contradiction again. This means that the common cause does not exist. This is an important example for two reasons. First, it shows that the strongest cause $b$ cannot be ``improved'' towards the common cause $C$. Second, it shows a concrete example, where the common cause does not exist. 

\section{Maximum likelihood and nonidentifiablity} 
\label{obs_non}

Recall the usual setup of parameter estimation \cite{cox_hinkley}. We want to estimate the unknown parameters $\theta$ in the probability $p_\theta(x )={\rm Pr}(X=x)$ of the random variable $X=\{x\}$. The estimation is done from a i.i.d. sample $(x_1, x_2, \dots, x_N)$. These parameters can be estimated by maximizing  
the likelihood function  over $\theta$ \cite{cox_hinkley}:
\be 
\label{3}
\mathcal{L}_N(\theta) = \frac{1}{N} \sum_{i=1}^N \ln p_\theta(x_i).
\ee
  According to the law of large numbers, for $N \gg 1$ the frequencies of various values in the sample coincide with the true probabilities, so we can rewrite  (\ref{3}) as 
\be 
\label{beta_rephrased}
\mathcal{L}_N(\theta) = \sum_{x} p_{\theta^*}(x)\ln p_\theta(x), 
\ee
where $\theta^*$ are the true parameters with which the sample was generated. 
The following relation obviously holds  
\be
\underset{\theta}{\arg \max}\sum_x p_{\theta^*}(x)\ln p_\theta(x) = \underset{\theta}{\arg \min}
\sum_x p_{\theta^*}(x)\ln \frac{ p_{\theta^*}(x) }{  p_\theta(x) }.
\ee
Since the relative entropy $\sum_x p_{\theta^*}(x)\ln \frac{ p_{\theta^*}(x) }{  p_\theta(x) }\geq 0$ nullifies
only for 
\be
\label{maximizator}
p_\theta(x) = p_{\theta^*}(x),~~{\rm for~~all}~~ x, 
\ee
the maximizer of (\ref{beta_rephrased}) holds (\ref{maximizator}). Now if (\ref{maximizator}) implies 
\be 
\label{eq_theta}
\theta = \theta^*,
\ee
the model is called identifiable \cite{cox_hinkley}. Otheriwse, when (\ref{maximizator}) holds, but (\ref{eq_theta}) does not, the model is called non-identifiable.

\section{Generalized likelihood and the free energy in statistical mechanics}
\label{free-energy}

Here we explain in which sense the generalized likelihood (\ref{beta0}) is similar to the (minus) free energy in statistical mechanics. 
This analogy is structural, i.e., it relates to the form of (\ref{beta0}), and not to applicability of any approximate method. We relate
$-\ln p(a,b,c)$ with the energy of a physical system, where $C$ and $(A,B)$ are respectively fast (hidden) and slow (observed) variables. Here fast and slow connect with (resp.) hidden and observed, which agrees with the set-up of statistical physics, where only a part of variables is observed \cite{free}. Then (\ref{beta0}) connects to the negative
nonequilibrium free energy with inverse temperature $\beta$ \cite{free}.  Here
nonequilibrium means that only one variable (i.e., $C$) is thermalized 
(i.e., its conditional probability is Gibbsian), while
the free energy has several physical meanings \cite{free}; e.g. it is a generating
function for calculating various averages and also the (physical) work
done under a slow change of suitable externally-driven parameters
\cite{free}. The maximization of (\ref{beta0}) naturally relates to the
physical tendency of decreasing free energy (one formulation of the
second law of thermodynamics) \cite{free}. 

Though formal, this correspondence with statistical physics can be
instrumental in interpreting $L_\beta$. E.g. we shall see that the
maximizer of $L_{\beta\leq 1}$ is unique (in contrast to maximizers of
$L_{\beta> 1}$), and this fact can be related to sufficiently high
temperatures that simplify the free energy landscape. Similarities between likelihood functions and free energy can also be used to apply powerful approximate methods of statistical physics to inference; see e.g. \cite{lamont}.

\section{No causes for independent events}
\label{anal1}

Consider the uncoupled situation: 
\BEA
\label{go0}
p(a,b)=p(a)p(b), 
\EEA
and assume that we look for 
the common cause by maximizig the generalized likelihood:
\BEA
\label{go1}
&&L_\beta=\frac{1}{\beta}\sum_{ab}p(a,b)\ln\sum_c p^\beta(c)p^\beta(a|c)p^\beta(b|c),\\
&&\sum_c p(c)p(a|c)p(b|c)=p(a,b).
\label{go2}
\EEA
Using (\ref{go0}, \ref{go2}) we find for (\ref{go1}):
\BEA
&&L_\beta= \frac{1}{\beta}\sum_{a,b}p(a,b)\ln\Big[ p^\beta(a,b) \sum_c p^\beta(c|a,b)  \Big]=\sum_{a,b}p(a,b)
\ln p(a,b) + \frac{1}{\beta}\sum_{a,b}p(a,b)
\ln\Big[ \sum_c p^\beta(c|a,b)  \Big]\\
&&=\sum_{a}p(a)\ln p(a)+\sum_{b}p(b)\ln p(b)\nonumber\\
&& +\frac{1}{\beta}
\sum_{ab}p(a)p(b)\ln\sum_c \pi^\beta(c|a,b),\\
&&\pi(c|a,b)=\frac{p(a,c)p(b|c)}{p(a,b)}=\frac{p(c|a)p(c|b)}{p(c)},
\label{go3}
\EEA
with constraint 
\BEA
\sum_c \pi(c|a,b)=1.
\label{go4}
\EEA
Note that constraint (\ref{go4}) does not depend on $(a,b)$. We now have:
\BEA
&&{\rm max}[L_\beta]=\sum_{a}p(a)\ln p(a)+\sum_{b}p(b)\ln p(b)\nonumber\\
\label{go5}
&& +\frac{1}{\beta}{\rm max}\left(
\sum_{ab}p(a)p(b)\ln\sum_c \pi^\beta(c|a,b)\right)\\
\label{goo5}
&&\leq 
\sum_{a}p(a)\ln p(a)+\sum_{b}p(b)\ln p(b)\\
&& +\frac{1}{\beta}
\sum_{ab}p(a)p(b)\,{\rm max}\left(\ln\sum_c \pi^\beta(c|a,b)\right).
\label{go6}
\EEA
In view of (\ref{go4}), the maximization in (\ref{go6}) is straightforward: 
$\ln\sum_c\pi^\beta(c|a,b)$ is a concave function of $\pi(c|a,b)$ for $\beta<1$, 
because it is made of concave functions. Hence the unique maximum of $\ln\sum_c\pi^\beta(c|a,b)$
is 
\BEA
\label{go7}
\pi(c|a,b)=\frac{1}{n},\qquad \sum_{c=1}^n \frac{1}{n}=1,
\EEA
which obviously saturates the inequality in (\ref{goo5}).

Eqs.~(\ref{go6}, \ref{go7}) imply:
\BEA
\label{go8}
&& p(c|a)=p(c|b)=p(c)=\frac{1}{n},\\
&& {\rm max}[L_\beta]=\sum_{a}p(a)\ln p(a)+\sum_{b}p(b)\ln p(b)+\frac{1-\beta}{\beta}\ln n,
\label{go9}
\EEA
where (\ref{go8}) means that the method predicts no cause, as it should.

The same method that led us to (\ref{go9}) can be employed for establishing generally not reachable upper bound on 
${\rm max}[L_\beta]$. Indeed, note that (\ref{go1}, \ref{go2}) can be written as 
\BEA
\label{go11}
&&L_\beta=\sum_{ab}p(a,b)\ln p(a,b)+
\frac{1}{\beta}\sum_{ab}p(a,b)\ln\sum_c \pi^\beta(c|a,b),\\
&& \pi(c|a,b)=p(c)p(a|c)p(b|c)/p(a,b),\qquad \sum_c \pi(c|a,b)=1.
\label{go12}
\EEA
Then (\ref{go11}) implies with the same method as for (\ref{go9}):
\BEA
\label{go13}
&&{\rm max}[L_\beta]\leq \sum_{ab}p(a,b)\ln p(a,b)+
\frac{1}{\beta}\sum_{ab}p(a,b)\,{\rm max}\left(\ln\sum_c \pi^\beta(c|a,b)\right),\\
&& = \sum_{a,b}p(a,b)\ln p(a,b)+\frac{1-\beta}{\beta}\ln n,
\label{go14}
\EEA
where the maxima in the last equation of (\ref{go13}) are reached for 
$\pi(c|a,b)=1/n$, which leads to $p(c)=\frac{1}{n}$. However, these two equations are consistent with the screening condition for $C$ only for $p(a,b)=p(a)p(b)$. This means that for $p(a,b)\not= p(a)p(b)$, (\ref{go14}) is an upper bound for ${\rm max}[L_\beta]$. 

Let us now generalize the above conclusion. Assume the following situation: 
\BEA
\label{go15}
&&L_\beta=\frac{1}{\beta}\sum_{a_1a_2b}p(a_1)p(a_2,b)\ln\sum_c p^\beta(a_1,a_2,c)p^\beta(b|c),\\
&&\sum_c p(a_1,a_2,c)p(b|c)=p(a_1)p(a_2,b).
\label{go16}
\EEA
Now (\ref{go15}) can be written as
\BEA
\label{go18}
&&L_\beta=\sum_{a_1}p(a_1)\ln p(a_1)+
\frac{1}{\beta}\sum_{a_1}p(a_1)\sum_{a_2b}p(a_2,b)\ln\sum_c p^\beta(a_2,c|a_1)p^\beta(b|c),\\
&&\sum_c p(a_2,c|a_1)p(b|c)=p(a_2,b).
\label{go19}
\EEA
Hence we find:
\BEA
&& {\rm max}[L_\beta]=\sum_{a_1}p(a_1)\ln p(a_1)+
\frac{1}{\beta}{\rm max}\left(\sum_{a_1}p(a_1)\sum_{a_2b}p(a_2,b)\ln\sum_c p^\beta(a_2,c|a_1)p^\beta(b|c)\right)\nonumber\\
&& \label{go20} \\
&&\leq 
\sum_{a_1}p(a_1)\ln p(a_1)+
\frac{1}{\beta}\sum_{a_1}p(a_1){\rm max}\left(\sum_{a_2b}p(a_2,b)\ln\sum_c p^\beta(a_2,c|a_1)p^\beta(b|c)\right).
\label{go21}
\EEA
Let us now recall (\ref{go16}) and write the conditional maximization in (\ref{go21}) as 
\BEA
{\rm max}\left(\sum_{a_2b}p(a_2,b)\ln\sum_c p^\beta(a_2,c|a_1)p^\beta(b|c)\right) \quad {\rm under}\quad
\sum_c p(a_2,c|a_1)p(b|c)=p(a_2,b).
\label{oru}
\EEA
It is now clear that $a_1$ in this conditional maximization is just a dummy index. 
Hence the maximum in (\ref{oru}) is reached for
\BEA
p(a_2,c|a_1)=p(a_2,c).
\label{go22}
\EEA
This also means that the inequality in (\ref{go21}) is satisfied. Hence the maximization in (\ref{go20}) is solved
via (\ref{go22}), where $p(a_2,c)$ is found from the restricted maximization:
\BEA
\label{go23}
&&
\frac{1}{\beta}{\rm max}\left(\sum_{a_2b}p(a_2,b)\ln\sum_c p^\beta(a_2,c)p^\beta(b|c)\right),\\
&&\sum_c p(a_2,c)p(b|c)=p(a_2,b).
\label{go24}
\EEA

\section{Triple information: reminder}
\label{tritri}

The triple information $I(\A_1:\A_2:\B)$ quantifies 
three-variable correlations present in  $p(a_1,a_2,b)$
\cite{triple,triple_f}:
\be
\label{aa11}
I(A_1:A_2:\B)&=&I(A_1:A_2)-I(A_1:A_2|\B)\\
\label{aa12}
&=& I(A_1:\B)+I(A_2:\B)-I(A_1A_2:\B)\\
&=&H(A_1)+H(A_2)+H(\B)-H(A_1A_2)-H(A_2\B)-H(A_1\B)+H(A_1A_2\B),
\label{aa13}
\ee
where $H(A_1 A_2\B)$ is the joint entropy of random variables $A_1$, $A_2$ and $\B$, and $I(A_1:\B)$ is the mutual information. $I(A_1:A_2:\B)$ is the change of the mutual information between any pair, e.g. $A_1$ and $A_2$, when introducing the context of $\B$; see (\ref{aa11}). Eq.~(\ref{aa13}) shows that $I(A_1:A_2:\B)$ has a set-theoretic meaning of triple overlap, and is invariant with respect to any permutation of the sequence $A_1, A_2, \B$.  Eq.~(\ref{aa12}) shows that $I(A_1:A_2:\B)$ also quantifies the sub-additivity of the mutual information with respect to joining any pair of the three random variables.

$I(A_1:A_2:\B)$ becomes zero when at least one random variable is independent from others, e.g., $p(a_1,a_2,b)=p(a_1,a_2)p(b)$. In contrast to two-variable correlations quantified by $I(A_1:A_2)\geq 0$ [$I(A_1:A_2)=0$ only when $A_1$ and $A_2$ are independent], $I(A_1:A_2:\B)$ can have either sign.  Now $I(A_1:A_2:\B)>0$ means the informations contained in the correlations $A_1:B$ and $A_2:B$ are [mutually] redundant, since when joining $A_1$ and $A_2$, the resulting mutual information decreases; cf.~(\ref{aa12}). Analogously, $I(A_1:A_2:\B)<0$ means that by joining $A_1$ and $A_2$ one creates information that was absent in $A_1:\B$ and $A_2:\B$ separately: the whole is more than the sum of its parts \cite{triple,triple_f,neural}. This scenario relates to the physical concept of frustration \cite{triple_f}, and finds applications in neuroscience \cite{neural}.

\section{Derivation of Eq.~(\ref{ut})}
\label{cond_ineq}

Consider the case when the joint probability of $A_1$, $A_2$ and $\B$ holds (\ref{guenter}). This situation can be modeled with $p(a,b) = p(a_2) p(a_1|a_2) p(b|a_2)$. Below we will prove that the most likely common cause (\ref{ccp3}) for this situation satisfies (\ref{ut}). Starting from (\ref{beta1}, \ref{bvana}) we note that in
\BEA 
\label{la1}
&& L_\beta = \frac{1}{\beta}
\sum_{a_1, a_2, b} p(a_1, a_2) p(b|a_2)  \ln{\sum_c p^\beta(a_1,a_2, c) p^\beta(b|c)} = \sum_{a_1, a_2} p(a_1, a_2) \ln{p(a_1, a_2)} \\ 
&&+ \frac{1}{\beta}  
\sum_{a_1, a_2, b} p(a_2) p(a_1 | a_2)p(b|a_2) \ln{\sum_c p^{\beta}(c|a_1, a_2) p^\beta(b|c)},
\label{lala}
\EEA
only (\ref{lala}) need to be optimized. Using the concavity of $L_{\beta<1}$ we have in (\ref{la1}):
\be 
\begin{split} \label{concineq}
&\sum_{a_2} p(a_2) \sum_{a_1} p(a_1 | a_2) \ln{\sum_c p^{\beta}(c|a_1, a_2) p^\beta(b|c)} \leq
\sum_{a_2} p(a_2) \ln{\sum_c (\sum_{a_1} p(a_1 | a_2) p(c|a_1, a_2))^\beta p^\beta(b|c)} \\&= \sum_{a_2} p(a_2) \ln{\sum_c p^\beta(c|a_2)  p^\beta(b|c)}.
\end{split}
\ee 
Now constraints (\ref{bvana}) together with (\ref{guenter}) can be written as
\be \label{appconst1}
p(b|a_2) = \sum_{c} p(c| a_1, a_2) p(b|c), \quad \sum_b p(b|c) =1.
\ee 
Assume that the maximization of (\ref{lala}) under constraints (\ref{appconst1})
resulted in $\hat{p}(c|a_1, a_2), \hat{p}(b| c)$. Obviously, the inequality (\ref{concineq}) holds also for the maximizers
\be 
\sum_{a_2} p(a_2) \sum_{a_1} p(a_1 | a_2) \ln{\sum_c \hat{p}^{\beta}(c|a_1, a_2) \hat{p}^\beta(b|c)} \leq
\sum_{a_2} p(a_2) \ln{\sum_c \hat{p}^\beta(c|a_2)  \hat{p}^\beta(b|c)}.
\label{sapo}
\ee 
On the other hand, we can maximize (\ref{lala}) with an additional constraint 
\be
p(c|a_1, a_2) = \hat{p}(c|a_2).
\label{ferr}
\ee 
Since (\ref{ferr}) is compatible with (\ref{appconst1}), it can only decrease the value of maximum:
\be 
\max_{p(c|a_1, a_2) p(b|c)}\left[{\frac{1}{\beta}  
\sum_{a_1, a_2, b} p(a_2) p(a_1 | a_2) \ln{\sum_c p^{\beta}(c|a_1, a_2) p^\beta(b|c)}}\right] \geq 
\frac{1}{\beta} \sum_{a_2, b} p(a_2) \ln{\sum_c \hat{p}^\beta(c|a_2)  \hat{p}^\beta(b|c)}.
\label{duda}
\ee
Comparing (\ref{duda}) with (\ref{sapo}) we get 
\be 
\max_{p(c|a_1, a_2) p(b|c)}\left[{\frac{1}{\beta}  
\sum_{a_1, a_2, b} p(a_2) p(a_1 | a_2) \ln{\sum_c p^{\beta}(c|a_1, a_2) p^\beta(b|c)}}\right] = 
\frac{1}{\beta} \sum_{a_2, b} p(a_2) \ln{\sum_c \hat{p}^\beta(c|a_2)  \hat{p}^\beta(b|c)}.
\ee
Thus, we deduced that the maximum is achieved for [cf.~(\ref{ut})]
\be \label{prvpc} 
\hat{p}(c|a_1, a_2)  = \hat{p}(c|a_2).  
\ee

\section{Minimization of the entropy $H(C)$ of $C$ for 3 binary variables $A$, $B$, and $C$} 
\label{ff}

Consider the common cause set-up (\ref{cccp}) for binary variables $A=\{a,\aa\}$, $B=\{b,\bb\}$, and $C=\{c,\cc\}$. Start from
\BEA
\label{u1}
&&p(a,b)=p(c)p(a|c)p(b|c)+[1-p(c)]p(a|\cc)p(b|\cc),\\
\label{u2}
&&p(a)=p(c)p(a|c)+[1-p(c)]p(a|\cc),\\
&&p(b)=p(c)p(b|c)+[1-p(c)]p(b|\cc).
\label{u3}
\EEA
Find $p(a|\cc)$ from (\ref{u2}) and $p(b|\cc)$ from (\ref{u3}) and put them into (\ref{u1}) to express $p(c)$ via unknown $p(a|c)$ and $p(b|c)$:
\BEA
\label{u04}
p(c)&=&\frac{p(a,b)-p(a)p(b)}{p(a|c)p(b|c)-p(b)p(a|c)-p(a)p(b|c)+p(a,b)}\\
&&\leq {\rm min}\Big[ \frac{p(a)}{p(a|c)}, \frac{1-p(a)}{1-p(a|c)}, \frac{p(b)}{p(b|c)}, 
\frac{1-p(b)}{1-p(b|c)}  \Big],
\label{u4}
\EEA
where the inequalities in (\ref{u4}) comes from $0\leq p(a|\cc)\leq 1$ and $0\leq p(b|\cc)\leq 1$.

To minimize $H(C)=-p(c)\ln p(c)-[1-p(c)]\ln [1-p(c)]$ means to maximize ${\rm max}[p(c),1-p(c)]$, i.e., to maximize $p(c)$, if $p(c)>1/2$, and to minimize it if $p(c)<1/2$. Hence, we shall 
find optimas (minimas or maximas) of $p(c)$, and then select the one with the smallest $H(C)$. 
As seen from (\ref{u04}), optimas of $p(c)$ as a function of $p(a|c)$ and $p(b|c)$ coincide with those of 
\BEA
\label{kon}
p(a|c)p(b|c)-p(b)p(a|c)-p(a)p(b|c)+p(a,b).
\EEA
The Hessian of (\ref{kon}) over $p(a|c)$ and $p(b|c)$ under $0<p(a|c)<1$ and $0<p(b|c)<1$ refers to a saddle point. Hence optimas of (\ref{kon}) are reached at the boundaries and amount to one of the following 4 solutions:
\BEA
\label{re1}
&&p(a|c)=p(b|c)=0,\\
\label{re2}
&&p(a|c)=p(b|c)=1,\\
\label{re3}
&&p(a|c)=1, \quad p(b|c)=0,\\
\label{re4}
&&p(a|c)=0, \quad p(b|c)=1.
\EEA
It is seen from (\ref{u04}) that (\ref{re1}) produces 
\BEA
\label{hh1}
p(c)=\frac{p(a,b)-p(a)p(b)}{p(a,b)},
\EEA
while (\ref{re2}) leads to a different solution with
\BEA
p(c)=\frac{p(\aa,\bb)-p(\aa)p(\bb)}{p(\aa,\bb)}.
\label{hh2}
\EEA
Hence (\ref{re1}, \ref{re2}) refer to the correlated situation:
\BEA
p(a,b)-p(a)p(b)=p(\aa,\bb)-p(\aa)p(\bb)=p(a,b)p(\aa,\bb)-p(a,\bb)p(\aa,b)>0. 
\label{umwelt}
\EEA
To find the actual solution under condition (\ref{umwelt}), we should select in (\ref{hh1}, \ref{hh2}), $p(c)$ that provides the smallest $H(C)=-p(c)\ln p(c)-[1-p(c)]\ln [1-p(c)]$.

Likewise, (\ref{re3}, \ref{re4}) refer to the anti-correlated situation, when the sign in 
(\ref{umwelt}) is inverted. It should be clear that (\ref{hh1}, \ref{hh2}) do not apply to the anti-correlated situation, because they predict $p(c)<0$. 

We emphasize that in all cases (\ref{re1}--\ref{re4}), constraint (\ref{u4}) holds automatically.

\end{document}